# Electric Field Effect Analysis of Thin PbTe films on high-ε $SrTiO_3$ Substrate

A. V. Butenko, R. Kahatabi[a], V. Sandomirsky, and Y. Schlesinger[*]

Bar-Ilan University, Department of Physics, Ramat-Gan 52900, Israel

and

A. Yu. Sipatov and V. V. Volubuev,

National Technical University KPI, 61002 Kharkov, Ukraine

## Abstract

Thin PbTe films (thickness 500 ÷ 600 Å), deposited on $SrTiO_3$, have been investigated by electric field effect (EFE). The high resistivity of such thin films warrants a high sensitivity of the EFE method. The $SrTiO_3$ substrate serves as the dielectric layer in the Gate-Dielectric-PbTe structure. Due to the large dielectric constant of $SrTiO_3$, particularly at low temperatures, the electric displacement $D$ in the film reaches the high value of $\sim 10^8$ V/cm, and the EFE introduced charge into the PbTe film amounts to ~ 8 μC/cm$^2$. The high $D$ permits to measure the EFE resistance and Hall constant over a wide region of $D$, revealing the characteristic features of their $D$-dependence. An appropriate theoretical model has been formulated, showing that, for such films, one can measure the dependence of the Fermi level on $D$. In fact, we demonstrate that shifting the Fermi level across the gap by varying $D$, the density-of-states of the in-gape states can be mapped out. Our results show, that the PbTe layers studied, possess a mobility gap exceeding the gap of bulk PbTe.

[*] Corresponding author: Y. Schlesinger, Bar-Ilan University, Department of Physics, Ramat-Gan 52900, Israel. E-mail: schlesy@mail.biu.ac.il

## 1. Introduction

In our previous reports [1, 2] we presented the results of our investigation of thin films (thickness $L$ smaller than, or comparable with, the Debye screening length $L_S$, $L_S$-films for short) of PbTe in the “Gate / mica / PbTe $L_S$-film” (MDS, Metal-Dielectric-Semiconductor) structure. We have demonstrated, that the electronic transport properties (resistance, Hall constant and Seebeck coefficient) of PbTe $L_S$-films deposited on mica substrate (without any particular preparation method or pre-treatment), are determined predominantly by the PbTe/mica interface states, and not by the bulk PbTe properties. In particular, we found that PbTe $L_S$-films on mica substrate have an unusually high resistance in comparison with the usually low-resistive bulk PbTe [3]. The motivation of the present work was to ascertain whether this is a unique property of the PbTe/mica system or a general property of $L_S$-films deposited on a substrate with a high concentration of interface states. Moreover, the high dielectric constant of $SrTiO_3$, especially at low temperatures, results in a dielectric displacement larger by about an order of magnitude when compared to the PbTe/mica system. This is of high importance when using the electric field effect (EFE) method, as this markedly extends the range of the EFE controlled properties by introducing a larger amount of charge carriers into the semiconductor film.

In a recently published article [4] an investigation of $L_S$-film of Si on $SiO_2$ has been reported. Though, the term “$L_S$-film” has not been used explicitly, the conclusions support our assertion that the transport properties of these films are determined largely by the interface states.

## 2. Experimental

The $n$-type PbTe films have been deposited on the polished (111) surface of monocrystalline dielectric $SrTiO_3$ (STO). The films were obtained by thermal evaporation in a vacuum of $10^{-7}$ Torr at a rate of 2-3 Å/s on a heated substrate (~350 C). The STO substrates have not been treated or cleaned particularly before the deposition, and the evaporated films have not been annealed. That was done intentionally, to obtain a high density of interface states at the PbTe/STO boundary. The thickness of the films was ~ 550 nm, the thickness of the STO substrate was 0.2 mm.

The measurements have been carried out in a closed cycle refrigerator within a temperature interval of 80 – 300 K. The temperature stability (using a Lakeshore 91C temperature controller) was 0.02 K at the higher temperature, and 0.002 K below 100 K.

The MDS structure has been prepared by evaporating a gold gate on the $SrTiO_3$ side opposite the PbTe-side (Fig. 1a). For the electric field effect measurements, a voltage $V_g$ ($-1100$ V $\leq V_g \leq +1100$ V) has been applied (using a Keithley 2410 Source/Meter) between the PbTe and the gate. The leakage current was less than $10^{-9}$ A.

The resistance and Hall constant were measured as a function of the EFE voltage and of temperature. The schematic drawing of the measurement system is shown in Fig. 1b. The measurements have been carried out in the four-point configuration. The ratio of the sample length to its width was $l/b$ = 2.83. The measuring current (using a Keithley 220 Current Source) was automatically adjusted to the changing resistance of the sample, to keep the power dissipation below $10^{-7}$ W. The resistance was measured by a Keithley 6514 electrometer. The Hall voltage was measured by a Keithley 182 NanoVoltmeter. The magnetic field was created by a variable field 6.5 kGauss electromagnet. The whole experimental process and data collection have been computer-controlled, using the ViewDac software package.

## 3. Analysis of the results

A general theoretical treatment of EFE for semiconductor films of arbitrary thickness consists in solving the Poisson equation with corresponding boundary conditions. The space charge in the Poisson equation is expressed through the energy spectrum of the semiconductor and the Fermi function. This theory has been described in detail and realized for the case of an intrinsic semiconductor layer in [5], and for the impurity semiconductor with interface states on the boundary of the dielectric/semiconductor in [6]. Technically, the problem involves the solution of a system of three transcendental equations, to determine the Fermi level ($E_F$), and the interface band bending ($\varphi_s$). The solution can be reached only numerically, for a given set of semiconductor parameters. Thus, a quantitative analysis of the theoretical predictions, and their comparison with experimental results, is rather complicated.

The theory can be markedly simplified for $L_S$-films, and, respectively, its results become much more transparent. It also turns out, that using this simplified theory, a richer amount of information on the PbTe-films properties can be extracted from the data.

The transition from the general case to $L_S$-film is presented schematically in Fig. 2. The arguments justifying this transition are the following:

(1) As the film thickness $L \lesssim L_S$, the variation of the charge concentration, across the thickness of the film, is small. Thus, the film can be considered to be homogeneously charged across the thickness $L$.

(2) The interface donors and acceptors affect the equilibrium carrier concentration and the occupation of bulk impurities within a region of thickness ~ $L_S$. Hence, one can express the interface surface impurity concentrations by the corresponding "bulk" volume concentrations, according to the rule

$$\frac{N_{ds}}{L} \to N_d;\ \frac{N_{as}}{L} \to N_a, \qquad (1)$$

where $N_{ds}$ and $N_{as}$ are the surface densities of interface donor and acceptor states, respectively, and $N_d$ and $N_a$ are the corresponding volume densities of the "bulk" states.

(3) The charge per unit area, introduced by EFE through the gate into the semiconductor film, is $D/4\pi$, where $D$ is the electric displacement in the dielectric. Since this charge is located in a layer region of thickness ~ $L_S$, and on the interface states, it can also be expressed by its "bulk" volume equivalent $Q = D/4\pi L$.

(4) The Poisson equation turns into the electroneutrality equation:

$$p + N_d^+ + N_e - n - N_a^- + N_p = -\frac{D}{4\pi e L}, \qquad (2)$$

where $p, n$ are the free hole and electron concentrations, $N_d^+$ and $N_a^-$ are the transformed concentrations of ionized donors and acceptors (former interface charged states), $N_e, N_p$ are the genuine, bulk PbTe, completely ionized, donor and acceptor concentrations, $e$ is the electronic charge. The sign of the right-hand part was chosen, consistently with our experiment, so that at $D < 0$ the positive charge of the film increases.

Choosing an energy spetrum, the left-hand part of Eq. (2) will depend only on the Fermi level, and Eq. (2) defines explicitly the form of the $D(E_F)$ function. Correspondingly, the inverse function $E_F(D)$ allows to find the carrier concentrations at given $D$ and $T$. Knowing the carrier mobilities, one can then calculate all the electron transport properties of the film.

The left-hand side terms of Eq. (2) are of very different orders of the magnitude. Therefore, the equation can be greatly simplified as follows:

(1) For PbTe at room temperature the value of $L_S \approx 500$ Å, and increases markedly with lowering temperature due to the increase of the dielectric constant. The bulk concentrations $N_e$, $N_p$ values are of the order of ~ $10^{17} - 10^{18}$ cm$^{-3}$. Hence, for the corresponding surface concentrations one obtains $(10^{17} - 10^{18}) \times L_S \approx 5\times(10^{11} - 10^{12})$ cm$^{-2}$. This value is smaller by two or three orders of magnitude than $N_{as}^{-}$, $N_{ds}^{+}$, as we have shown [1] for the mica/PbTe interface. It is reasonable to assume, that this will be so in the $SrTiO_3$/PbTe case as well (this statement will be actually confirmed later in this article). Then, one can neglect $N_e$, $N_p$ in Eq. (2). Thus, according to Eq. (2), the type of conductivity of the $L_S$-film is determined by the nature of the interface trapping states only, and not by the bulk impurities. In fact, in our case, the starting bulk material, from which the film was deposited, was usually of *n*-type, while almost all the films exhibit a *p*-type conductivity.

(2) The experiment shows that the electroconductivity depends strongly on temperature. Hence, the carrier concentrations, $p$ or $n$ depend on temperature. Then, they must be described by the Boltzmann formula for a non-degenerate electron gas. Further, our estimates show that the hole and the electron concentrations are much smaller than the charge introduced by EFE. (The graphs for $p(D,T)$ and $n(D,T)$ are presented below). Hence, in the left part of Eq. (2) it is sufficient to keep $N_d^{+}$ and $N_a^{-}$ only.

Thus, Eq. (2) assumes the form

$$N_d^{+} - N_a^{-} = -\frac{D}{4\pi eL}, \tag{3}$$

where

$$N_d^+ = \int_{E_v}^{E_c} N_d(E)\cdot\left[1 - f\left(\frac{E-E_F}{kT}\right)\right]dE;\; N_a^- = \int_{E_v}^{E_c} N_a(E)\cdot f\left(\frac{E-E_F}{kT}\right)dE$$

$$f\left(\frac{E-E_F}{kT}\right) = \frac{1}{1+\exp\left(\frac{E-E_F}{kT}\right)}. \tag{4}$$

$N_d(E)$ and $N_a(E)$ are the donor and acceptor density of states, $E_v$ and $E_c$ are the edges of the valence and conductance bands, respectively, or , as we argue below, the edges of a mobility gap.

It is rather important to stress that in the $L_S$-film the band bending is absent. Therefore, the band edges are parallel to $E_F$. Hence, $N_d^+, N_a^-$ as well as the carrier concentrations, do not depend on the coordinates across the film.

Thus, the EFE in a $L_S$-film causes the $E_F$ to shift with $D$ in accord with Eq. (2), thus changing $N_d^+$ and $N_a^-$. The hole and the electron concentrations vary by the shifting Fermi level,

$$p = N_v \cdot \exp\left(-\frac{E_F - E_v}{kT}\right);\quad n = N_c \cdot \exp\left(-\frac{E_c - E_F}{kT}\right);$$

$$N_{v,c} = 8\left(\frac{m_{p,n}\cdot kT}{2\pi\hbar^2}\right)^{3/2}. \tag{5}$$

Here, the density-of-states, $N_{v,c}$ includes the factor 8, which is specific for PbTe, $m_{p,n}$ are the hole and electron effective masses, respectively.

The temperature dependence of the $SrTiO_3$ dielectric constant [7], was used to calculate the electric displacement in our experiment at a given $V_g$ and $T$. An example of such graph, at $V_g$ = 1000 V, is shown in Fig. 3.

Figs. 4-6 show the measured dependences of the resistance $R$ and Hall constant $R_H$ on the electrical displacement and on the temperature, for a 550 Å thick PbTe film. Even though the resistance and the Hall constant have been measured over a wide range of temperatures (from 300 K down to 85 K), only in the low-temperature region the range of $D$ is wide enough to show the complete curves with their characteristic features.

Fig. 4 shows the dependence of the resistance on the electric displacement at different temperatures. These graphs exhibit the characteristic shape for EFE, with the resistance maximum at certain $D$. Since the maximum appears at a positive displacement, $D > 0$, where the PbTe layer assumes a negative charge, the conductivity of the original layer is of $p$-type.

The physical significance of the $R_H$-curves (Fig. 5) can be explained easily qualitatively. The left and right $D$-tails correspond to the hole-rich and to the electron-rich regions, respectively. The decrease of $|R_H|$ with $D$ in these regions is expected, according to the dependence of $|R_H|$ on the single type carrier concentrations, namely, $\propto 1/p$ and $1/n$ respectively. The in-between region, where $R_H$ changes sign, is characterized by bipolar conductivity. This curve was derived in detail in [5]. Qualitatively, this curve can be obtained by using the standard expression for $R_H$ together with Eq. (5), with $E_F$ varying over the gap as a function of $D$. It should be emphasized that such curve can be observed only in an $L_S$-film, where the carrier concentration change, due to the EFE, is approximately homogeneous over the entire film volume. In a thick layer, where the sign of $R_H$ is determined by the bulk properties, such curve cannot be obtained. In fact, observation of an $R_H$-curve, of the form shown in detail in Figs. 5 - 6, indicates clearly that the film under investigation is an $L_S$-film.

The point where $R_H = 0$ is located in the electron tail region. That confirms also that the original film has $p$-conductivity.

Since we deal here with an $L_S$-size layer, one can use for $R_H$, in the $p$- and $n$-tail regions, the corresponding formulas for single type carriers, and find the concentrations $p(D)$ and $n(D)$. These are shown in Fig. 7 at $T$=85 K and $T$=105 K. Since, the conductivity and the $R_H$ are known, the hole and electron mobilities, in the hole and electron tails, can be determined, see Fig. 8.

Comparing the EFE introduced charge $Q$ (Fig.3) and the values of $p$ and $n$, it is apparent that the latter are much smaller then $Q$. This justifies our assertion about the dominance of the $N_d^+$ and $N_a^-$ terms in Eq. (2).

The electron mobility ($\mu$ ~ 10,000 cm$^2$/V·s), as derived from these experimental results, was found to be much higher than the hole mobility ($\nu$ ~ 50 cm$^2$/V·s). Recalling the expression for $R_H$ in a bipolar conductor

$$R_H \propto \frac{p\nu^2 - n\mu^2}{\left(p\nu + n\mu\right)^2},$$

$R_H = 0$ when $\nu^2 p = \mu^2 n$. Fig. 5 shows, that this point is at the far side of the electron tail (the side of smaller $D$). From the value of $\mu/\nu$, it follows that the hole concentration here is still larger than the electron concentration by about six orders of magnitude. The Hall constant becomes negative because of the large electron mobility, and not due to the electron concentration exceeding the hole concentration.

Now, having obtained $p$ and $n$ and using Eq. (5), one can find the location of the Fermi level, $E_F$-$E_v$ and $E_c$-$E_F$ as a function of $D$ at given $T$, in the hole and electron tails. These graphs are given in Figs. 9a and 9b. An important conclusion ensues from these curves. It is known [8] that the temperature variation of the PbTe gap can be described by the function

$$E_g(T)\left[\text{meV}\right] = 190 + \frac{0.45 \times T^2}{T + 50}$$

Thus, at $T \approx 100$ K, $E_g(100)$ is ~220 meV. On other hand, Figs. 9a and 9b show that, at T=105 K, the gap (sum of $E_F$-$E_v$ and $E_c$-$E_F$) can be larger than 280 meV. This means that the PbTe $L_S$ –film on $SrTiO_3$ has a mobility gap exceeding $E_g$. Thus, $E_v$ and $E_c$ must be interpreted as the lower and upper edges of a mobility gap, respectively.

Now, we have to choose a reasonable energy spectrum of the $N_a(E)$ and $N_d(E)$ states. Since the film conductivity is $p$-type, the acceptor concentration exceeds the donor concentration. It is plausible to suggest that the acceptor states are located in the lower half of the gap, and the donor states in the upper one. Such distribution of states has been observed also in the mica/PbTe system [2].

Several different energy spectra have been examined to calculate $R_H(E)$ and to compare it to the experimental Hall curve as shown in Fig.5: (1) Monoenergetic acceptor and donor levels; (2) $N_a$(E) =const < $N_d(E)$ = const, $E_v < E < E_c$; (3) $N_a(E)$ =const, $E_v < E < E_g/2$, and $N_a(E)$ =0, $E_g/2 < E < E_c$; $N_d(E)$ =0, $E_v < E < E_g/2$ and $N_d(E)$ =const, $E_g/2 < E < E_c$.

All these schemes fail to reproduce satisfactorily the experimental results. On the other hand, we show in the following, that assuming a non-homogeneous distribution of the acceptor and donor levels in the lower and upper parts of the gap, respectively, result in a satisfactory agreement with the experimental results. We show next, that using this model one can derive the density of states $N_a(E)$ and $N_d(E)$ close to the lower and upper edges of the mobility gap. In fact, Figs. 9a and 9b show, that when the Fermi level resides in the lower part of the mobility gap, the hole concentration changes, evidently due to an interchange of holes between the acceptor levels and the valence band. Thus, in this region of $D$, on the left-hand side of Eq. (3), only $N_a^-$ varies. In the opposite case, in the electron region, when $E_F$ is shifted to the upper part of the mobility gap, only $N_d^+$ varies.

Let us differentiate the left and right-hand sides of Eq. (3) with respect to $E_F$ in the hole region. We obtain

$$\int_{E_v}^{E_c} N_a(E) \cdot \frac{\partial}{\partial E_F} f\left(\frac{E - E_F}{kT}\right) = \frac{1}{4\pi e L} \frac{\partial D}{\partial E_F}. \tag{6}$$

The right-hand part in Eq. (6) is known, having found $D(E_F)$ experimentally (see Figs. 9a and 9b). In the left-hand part of Eq. (6) the derivative of the Fermi function at low temperature is close to the $\delta$-function. Calculating asymptotically the integral on the left side, we obtain

$$N_a\left(E_F\right) = \frac{1}{\sqrt{2\pi} \cdot \pi \cdot e \cdot L} \cdot \frac{\partial D}{\partial E_F}. \tag{7}$$

The analogous formula for $N_d(E)$ in the electron region is obtained similarly.

Thus, by shifting $E_F$ by $D$, we probe the local density of states. To express more clearly its physical meaning, Eq. (7) can be written in the form

$$\frac{\partial E_F}{\partial D} = \frac{1}{\sqrt{2\pi}\pi e L} \cdot \frac{1}{N_a(E_F)}. \tag{7.1}$$

Indeed, the extent of shift of $E_F$ with $D$ is larger, the smaller the density-of-states, and it gets smaller with an increase of the density of states. The Fermi level will be pinned by a large density of states, as long as they are vacant. In the opposite case, in absence of local states, the Fermi level will shift with $D$ throughout the energy gap.

The density of states, derived by the method outlined above, is presented in Fig. 10. This spectrum changes slightly with temperature, but preserves its shape. The "acceptor" region, extending over a wide region of energy, contains a large number of experimental points, and therefore can be analyzed thoroughly. The "donor" region is very narrow, containing only a few measured points, making it difficult to extract any reliable information.

The next step is to substitute the functions $N_a(E)$ into Eq. (3) in the hole region, and $N_d(E)$ in the electron region, to obtain the corresponding function $E_F(D)$. This will, then, be compared with the experimental curves of $E_F(D)$, shown in Figs. 9a and 9b. To illustrate the method, the whole procedure will be carried out here in detail for $N_a(E)$.

Step (1): The experimental density of states was approximated by the function

$$N_{ai}(E) = 4.036\cdot 10^{21}\exp(-56.6E)\Theta(0.056-E)+ \\ +8.905\cdot 10^{20}\exp(-30.12E)\Theta(E-0.0561)\Theta(0.072-E)+ \\ +3.413\cdot 10^{20}\exp(-16.802E)\Theta(E-0.0721)\;\left[cm^{-3}\cdot eV^{-1}\right], \tag{8}$$

where $\Theta(x)$ is the Heaviside function $(\Theta(x<0)=0, \Theta(x\geq 0)=1)$. The energy in (8) is expressed in eV.

Step(2): The function $N_{ai}(E)$, (Eq.(8)) is substituted into Eq. (3). We obtain

$$N_{dt} - \int_{E_v}^{E_c} N_{ai}(E)\cdot f(E-E_F)dE = -\frac{D}{4\pi eL};\; N_{dt} = \int_{E_v}^{E_c} N_d(E)dE, \tag{9}$$

where $N_{dt}$ [$cm^{-3}$] is the total donor concentration. The Fermi level is below the donor states, which are totally ionized. Thus, Eq. (9) defines the function $D(E_F)$ up to the unknown constant $N_{dt}$.

The total acceptor concentration can be estimated by evaluating the integral $\int_0^{E_c} N_{ai}(E)dE$ with an upper limit $E_c$ or, in any case, larger than the upper boundary of the hole tail. According to Fig. 9a this is ~ 120 meV. The mobility gap at 105 K and 110 K is ~ 300 meV. Using the function, Eq. (8), it turns out that this integral depends only weakly on the upper limit. The integral varies from $7.4\times 10^{19}$ to $7.6\times 10^{19}$ $cm^{-3}$, when the upper limit is changed from 120 meV to 300 meV.

Step (3): For the sake of simplicity we will use the Fermi function, $f(E- E_F)$ at $T$=0 K. Then, the upper limit in the integral at the left-hand side of Eq. (9) is $E_F$. The constant $N_{dt}$ can be determined by substituting a pair of ($D$, $E_F$) from the experimental data (Fig. 9). This way one finds that $N_{dt}$ =7.08×10$^{19}$ cm$^{-3}$. Finally one obtains the function of $D(E_F)$

$$\frac{D(E_F)}{10^6\,\mathrm{V/cm}} = 7.04\times10^2\left[\frac{1}{7.08\cdot10^{19}}\int_0^{E_F-E_v} N_{ai}(E)\,dE - 1\right], \tag{10}$$

where ($E_F$ -$E_v$) is expressed in eV.

The calculated inverse function $E_F(D)$ is presented in Fig. 12a (curve 1) together with the experimental graph (curve 2) adopted from Fig. 9. The $R_H$ was then derived based on the calculated $E_F(D)$, shown in Fig. 12b (curve 1), and compared to the experimental data (curve 2). Taking into account the approximations made at deriving Eq. (10), the comparison is evidently quite satisfactory.

Since, the $D$ dependence of $E_F(D)$ in the region of the hole-tail coincides satisfactorily with the experimental data, so will also the resistance and the Hall constant.

The donor density-of-states (Fig.11) has been observed in a very narrow energy band, of ~ 10 meV, only. Moreover, this energy band is located considerably below (by ~ 150 meV) the upper edge of the mobility gap. Therefore, practically, it is difficult to extract from it any useful information.

The total concentrations $N_a$ and $N_d$ can be estimated also from Eq. (3), and Figs. 9a and 9b. According to Eq. (3)

$$\Delta N_a^- = \frac{\Delta D_p}{4\pi eL} \quad \text{and} \quad \Delta N_d^+ = \frac{\Delta D_n}{4\pi eL}, \tag{11}$$

where $\Delta N_a^-$, $\Delta N_d^+$ and $\Delta D_{p,n}$ are the increments of occupations and of the electric displacement in the $p$- and $n$-tails, respectively. As $\Delta D_p \approx 100\times10^6$ V/cm, then $\Delta N_a^- \approx 10^{19}$ cm$^{-3}$. Also, $\Delta D_n \approx 40\times10^6$ V/cm, thus $\Delta N_d^+ \approx 4\times10^{18}$ cm$^{-3}$ . Hence, $N_a \geq 10^{19}$ cm$^{-3}$, $N_d \geq 4\times10^{18}$ cm$^{-3}$. These values are consistent with our above estimates.

## 4. Summary

PbTe films, having a thickness of the order of magnitude of the Debye screening length ($L_S \approx 500 \div 600$ Å), deposited on $SrTiO_3$, have been investigated by electric field effect method. Due to the trapping of the charge carriers on the PbTe/$SrTiO_3$ interface states, such films acquire an unusually high resistivity. This appears to be a characteristic of such films, in the case of high enough concentration of interface defect states. The high dielectric constant of $SrTiO_3$ at low temperatures, results in a high value of the displacement $D$ (~ $10^8$ V/cm), and a correspondingly large amount of EFE introduced charge into the PbTe film (~ 8 μC/cm$^2$). The large $D$ permits to measure the EFE controlled resistance and Hall constant over a wide region of $D$, revealing the characteristic features of their $D$-dependence. We demonstrated also, that these transport properties are determined mainly by the nature of the interface and not by the bulk characteristics.

A theoretical model has been formulated specifically for the $L_S$ – films. Using this model, one can extract the dependence of the Fermi level on $D$ from the experimental data. We have demonstrated that shifting the Fermi level across the forbidden gap by varying $D$, the density-of-states of the in-gape states can be mapped out. Our results show also, that the PbTe layers studied, possess a mobility gap exceeding the gap width of bulk PbTe. This is apparently also a general feature of the $L_S$ – films.

**Figure captions**

**Fig. 1.** (a) The MDS structure

$L$, $d$ are thicknesses of PbTe and $SrTiO_3$, respectively; $V_g$ is the gate voltage.

$d = 0.02$ mm; $L = 55$ nm.

(b) The measurement scheme.

The resistance is determined by measuring $V_R$, and the Hall constant is determined by measuring $V_H$ .

**Fig. 2.** The transition from surface density to the corresponding ,"bulk", volume density in an $L_S$-homogeneous layer.

**Fig. 3.** The temperature dependence of the $SrTiO_3$ dielectric constant, $\varepsilon(T)$; the calculated temperature dependences of the electric displacement $D$ in $SrTiO_3$**,** and the "space charge" $D/4\pi{\cdot}e{\cdot}L$, (see text) at $V_g = 1000$ V.

**Fig. 4.** The dependence of the resistance on $D$ and $T$.

(a)$T = 140$ K; (b)$T$=120 K; (c)$T$=105 K; (d).$T$=85 K

**Fig. 5** The dependence of the Hall constant on $D$ and $T$. The logarithmic scale on the left corresponds to the positive $R_H$ values.

(a)$T = 140$ K; (b)$T$=120 K; (c)$T$=105 K; (d).$T$=85 K

**Fig. 6.** The dependence of the Hall constant on $D$ at 110 K (expanded scale).

The region of transition from the positive $R_H$ to negative $R_H$ is shown. The upper scale shows the corresponding space charge.

**Fig. 7.** The dependence of the carrier concentrations on the electric displacement. $T$ = 105 K and 85 K.

**Fig. 8.** The dependence of the carrier mobilities on the electric displacement. $T$ = 105 K and 85 K.

**Fig. 9.** The dependence of $E_F$ on $D$.

(a) the hole-tail; (b) the electron-tail.

**Fig. 10.** The acceptor density of states at different temperatures.

**Fig. 11.** The and donor density-of-state of the acceptors (left) and of the donors (right) at $T$ = 110 K.

**Fig. 12.** Comparison of calculated and experimental $E_F(D)$ and $R_H(D)$ in the hole-tail region.

(a) $E_F(D$, 105K), 1 –calculated from Eq. (10), 2 - experimental

(b) $R_H(D$, 105K) 1 –calculated from Eq. (4.1), 2 – experimental.

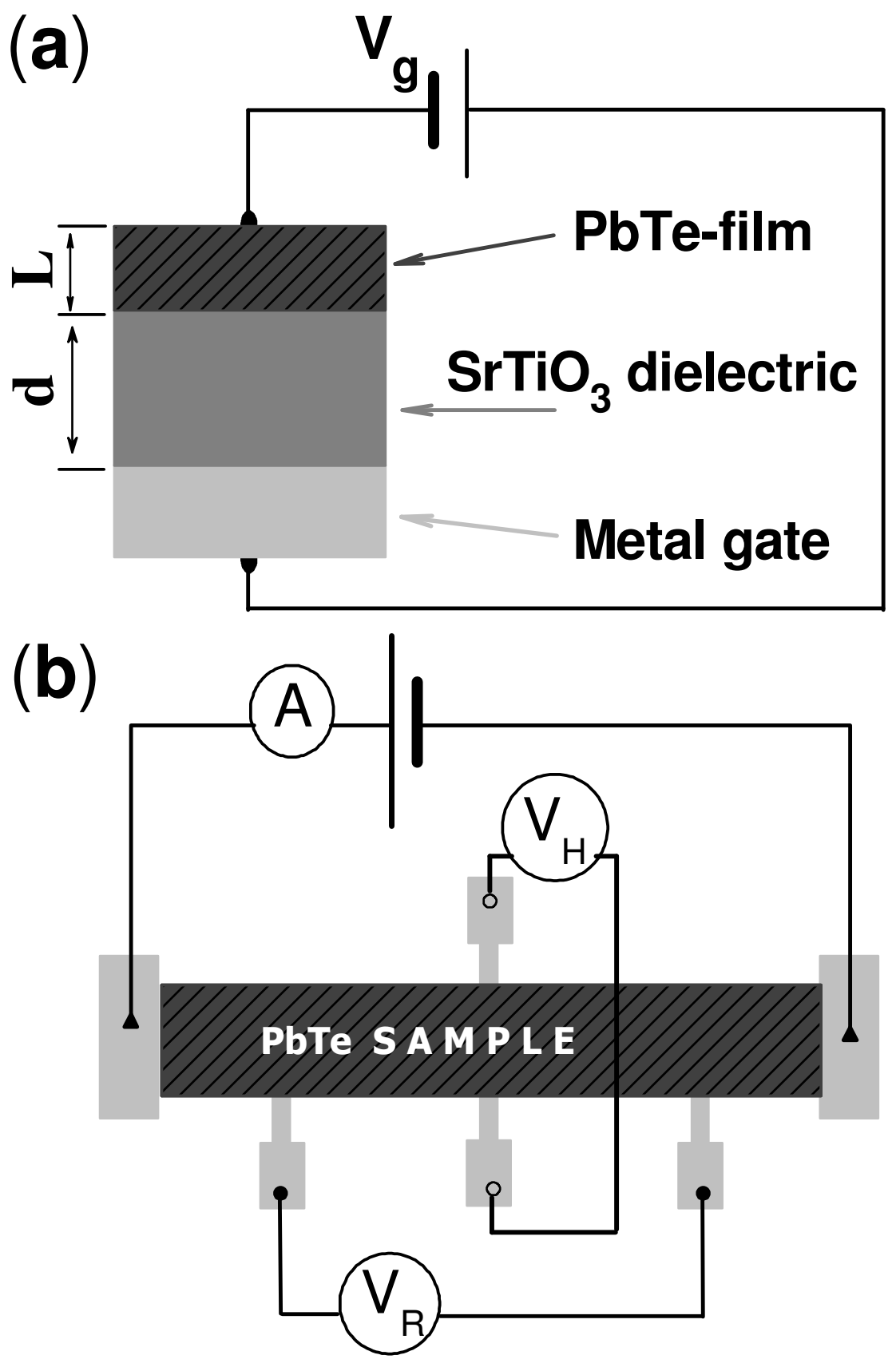

**Fig. 1**

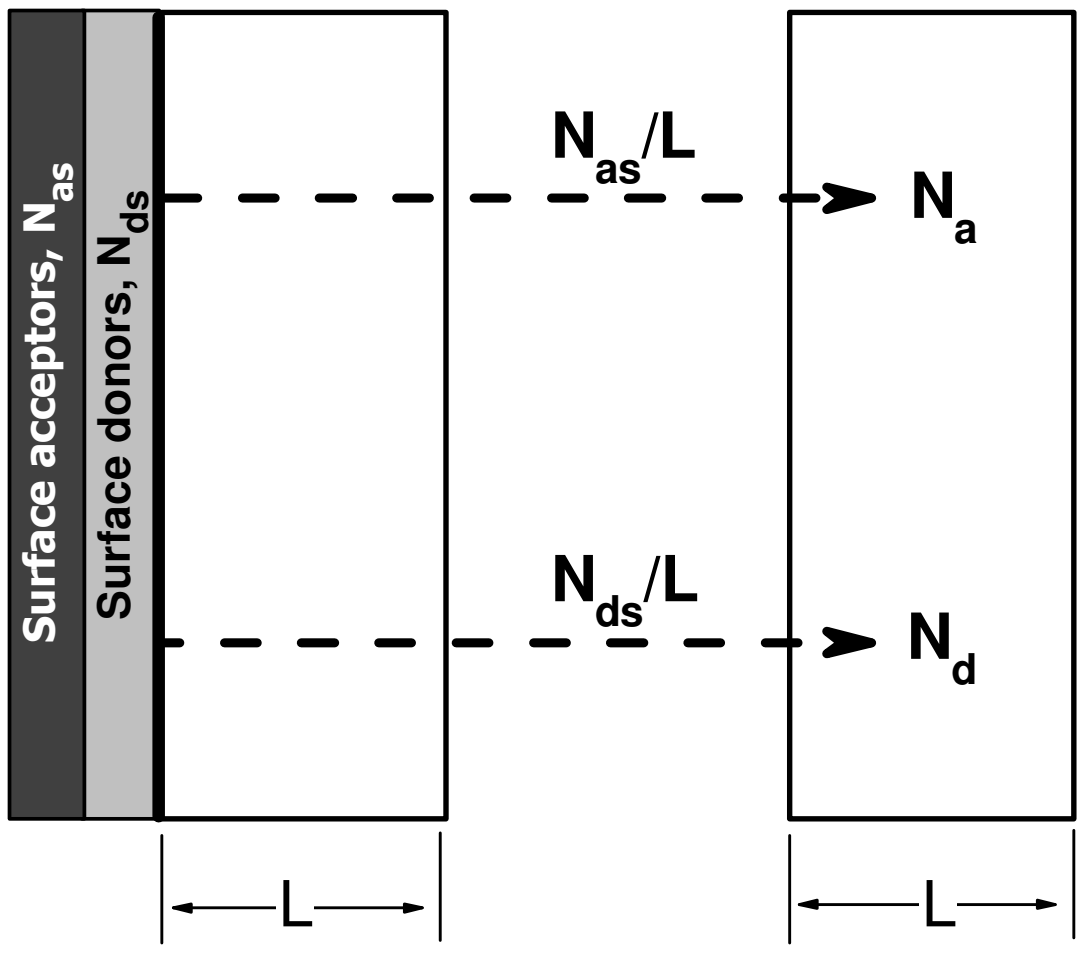

**Fig. 2.**

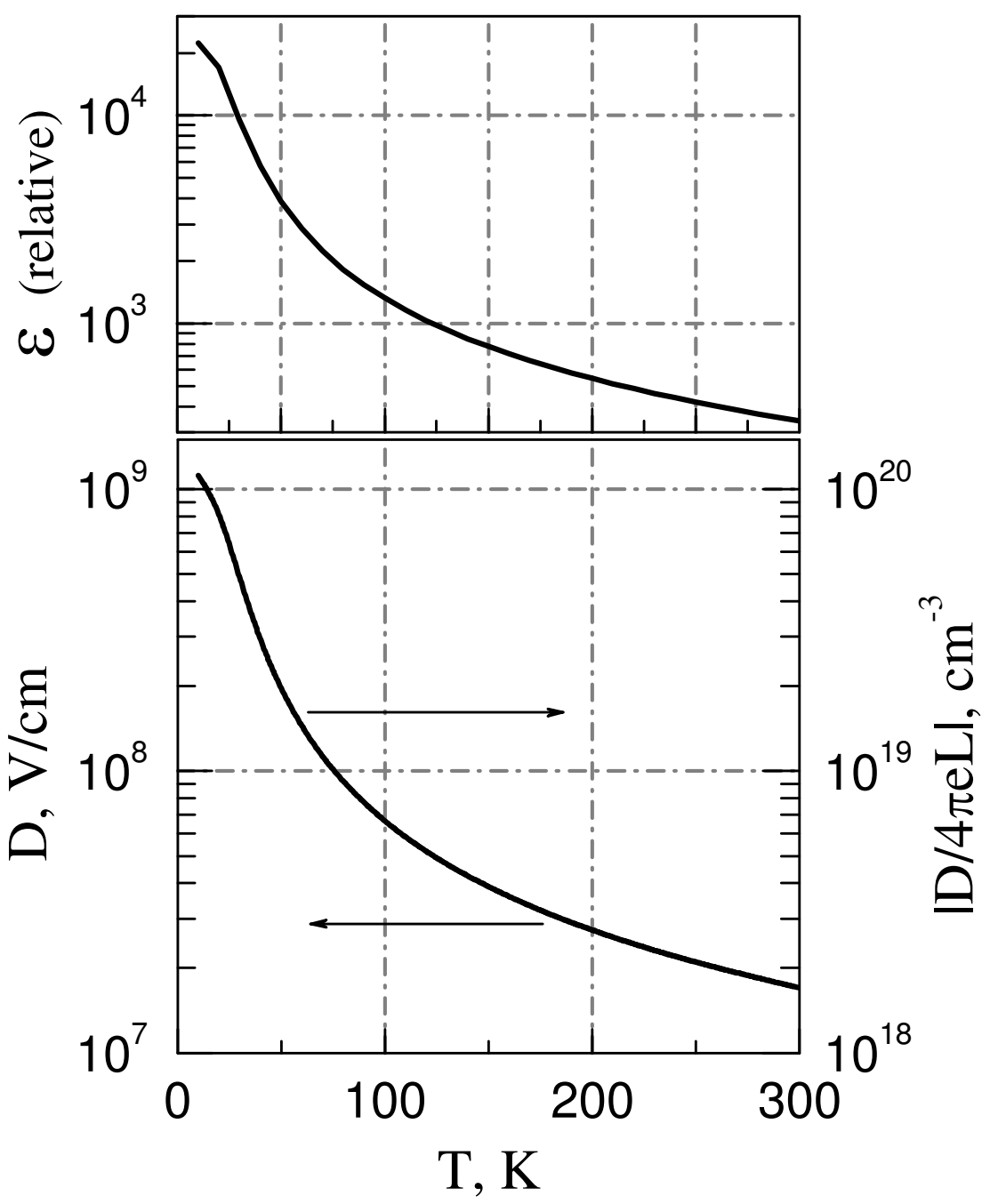

**Fig. 3.**

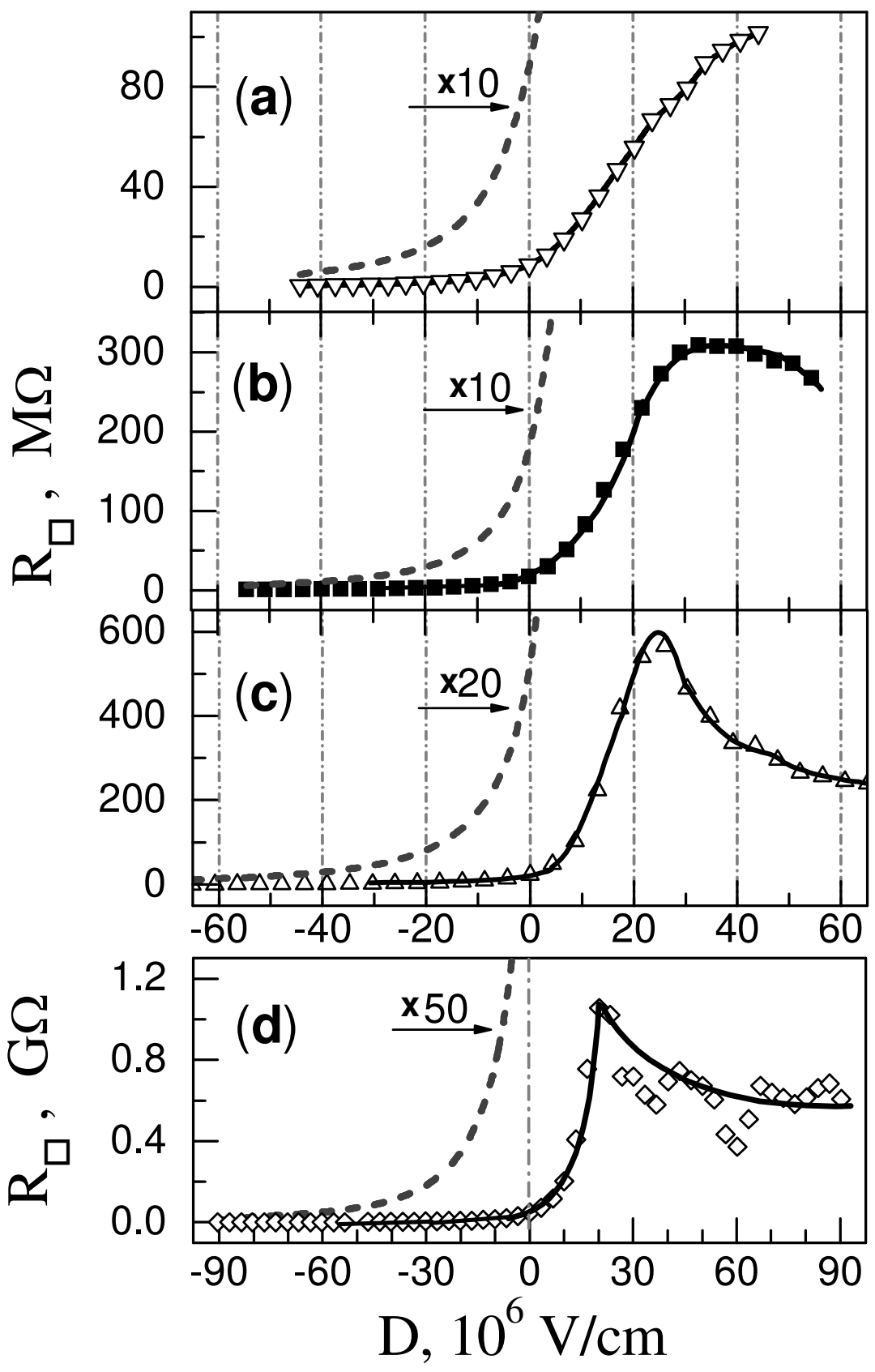

**Fig. 4.**

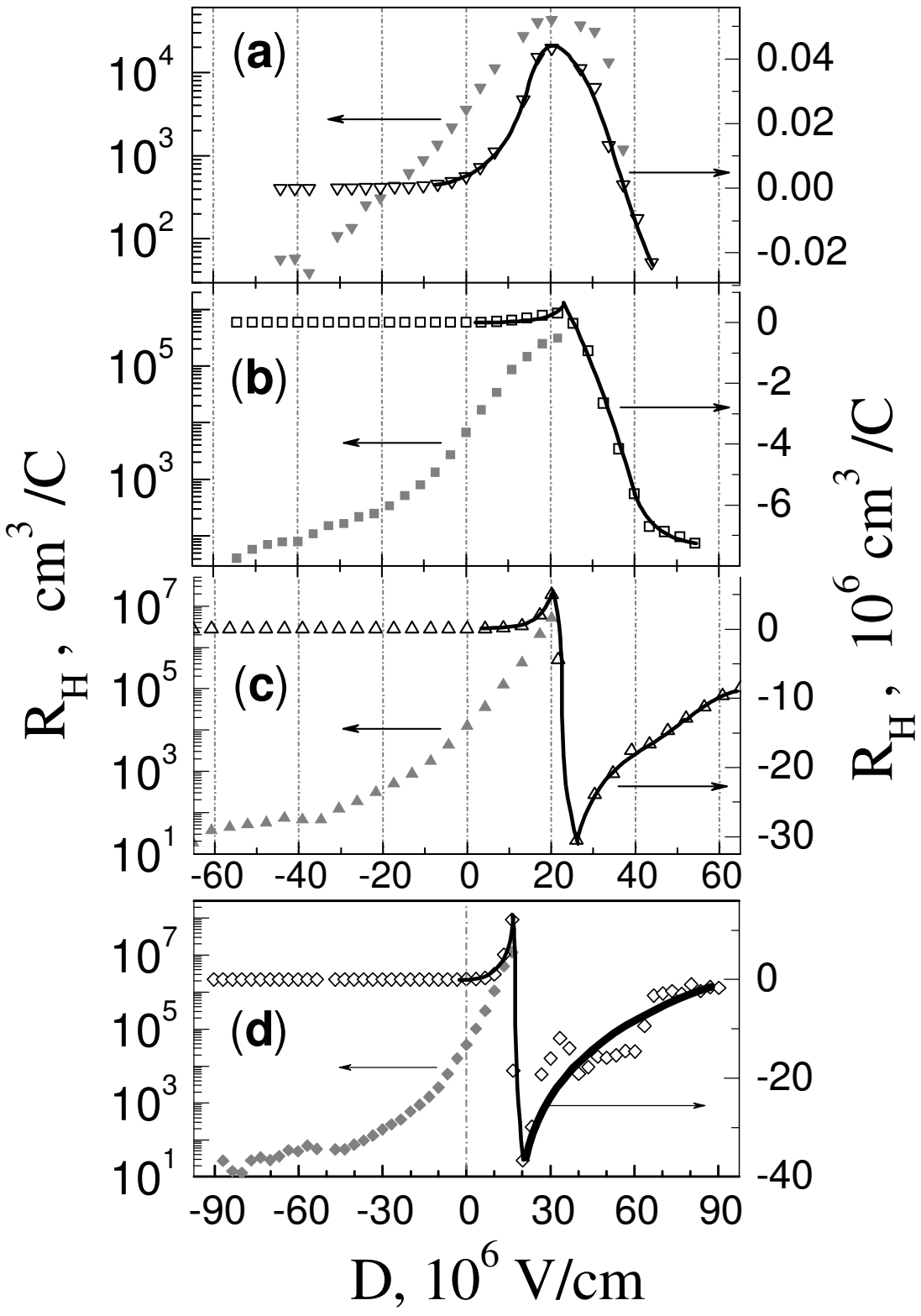

Fig. 5

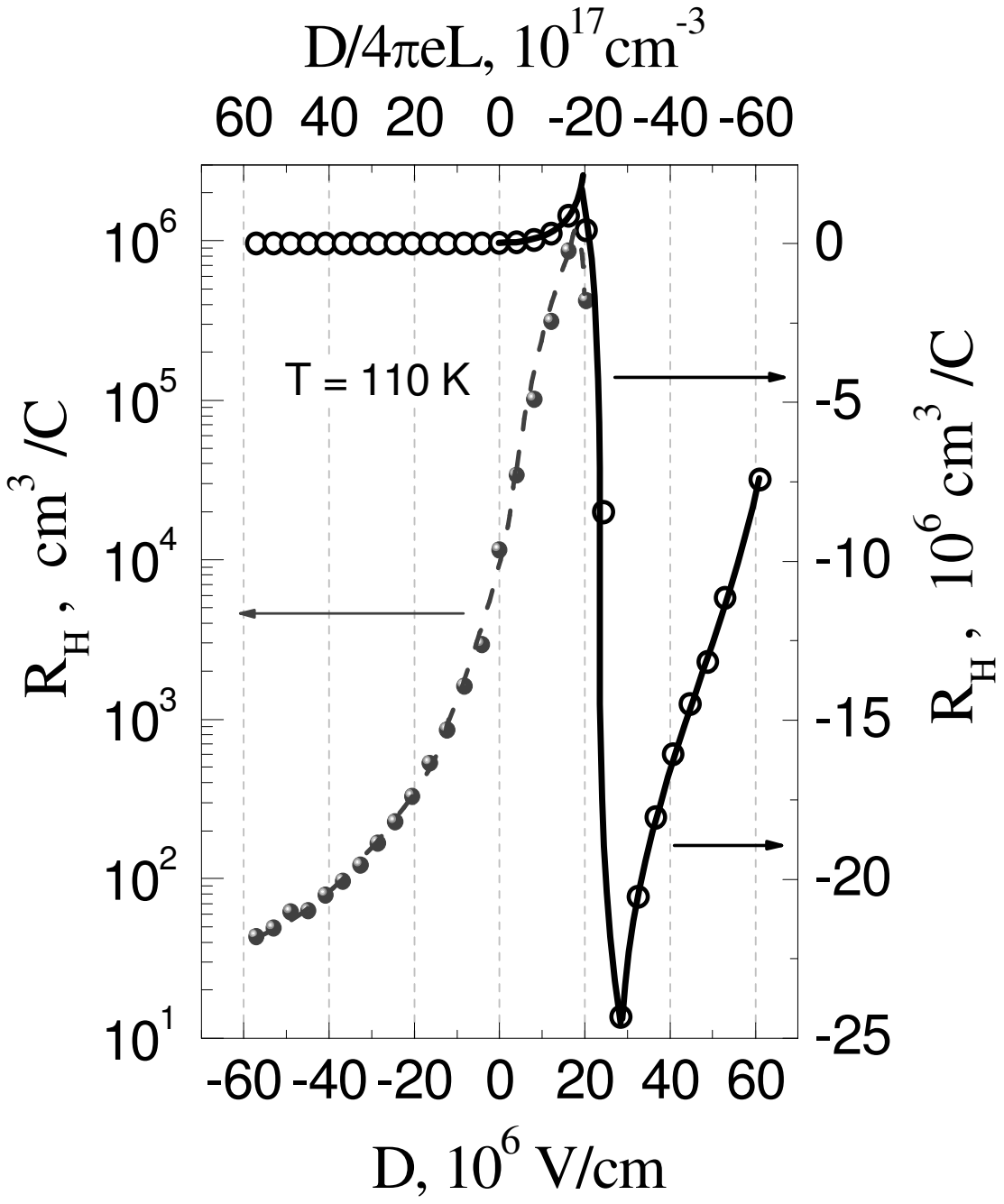
D/4πeL, $10^{17}$cm$^{-3}$
60 40 20 0 -20 -40 -60
$10^6$
$10^5$
$10^4$
$10^3$
$10^2$
$10^1$
$R_H$ , cm$^3$ /C
T = 110 K
0
-5
-10
-15
-20
-25
$R_H$ , $10^6$ cm$^3$ /C
-60 -40 -20 0 20 40 60
D, $10^6$ V/cm

**Fig. 6.**

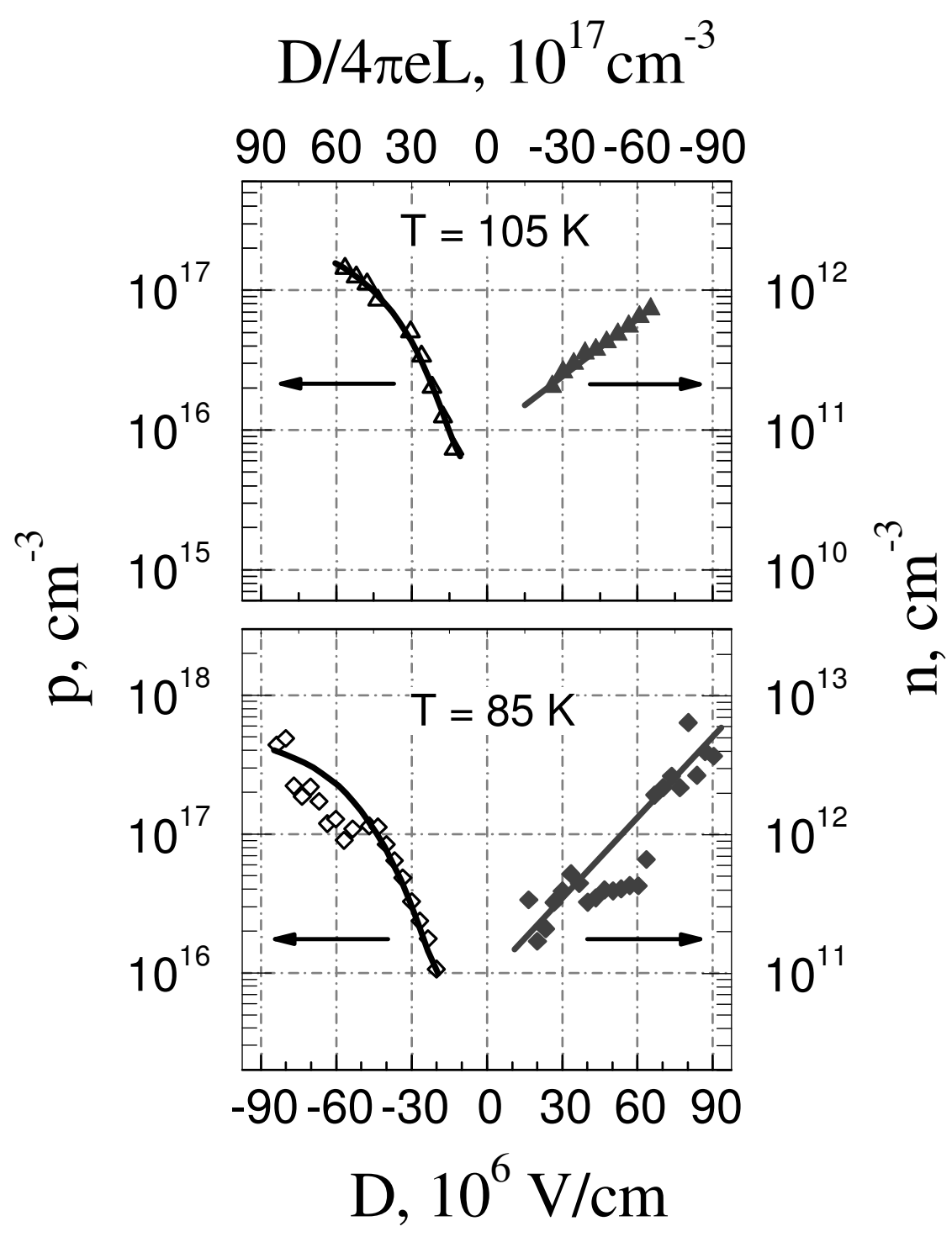

**Fig. 7.**

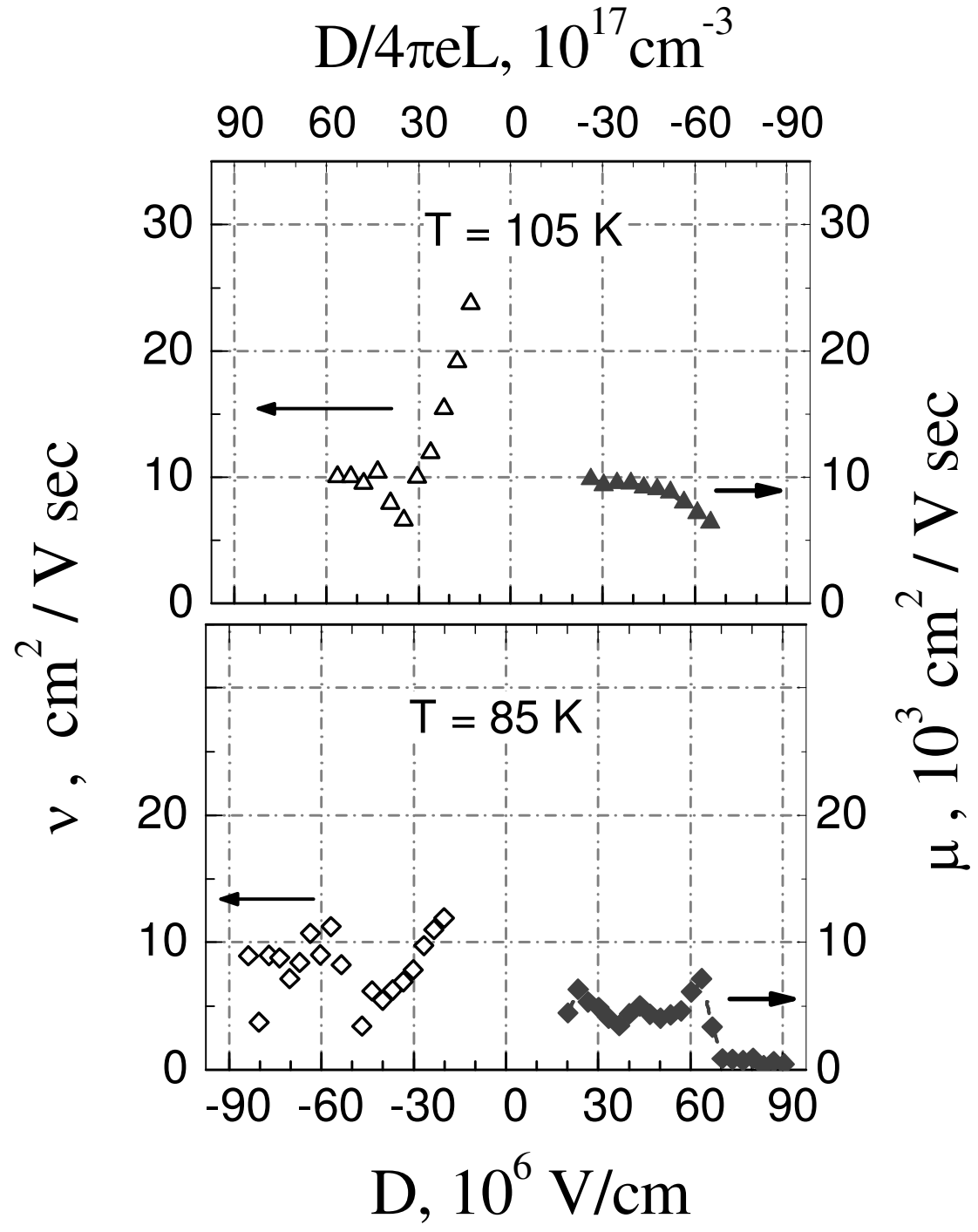
D/4πeL, $10^{17}$cm$^{-3}$
90 60 30 0 -30 -60 -90
T = 105 K
T = 85 K
ν , cm$^2$ / V sec
μ , $10^3$ cm$^2$ / V sec
0 10 20 30
-90 -60 -30 0 30 60 90
D, $10^6$ V/cm

Fig. 8.

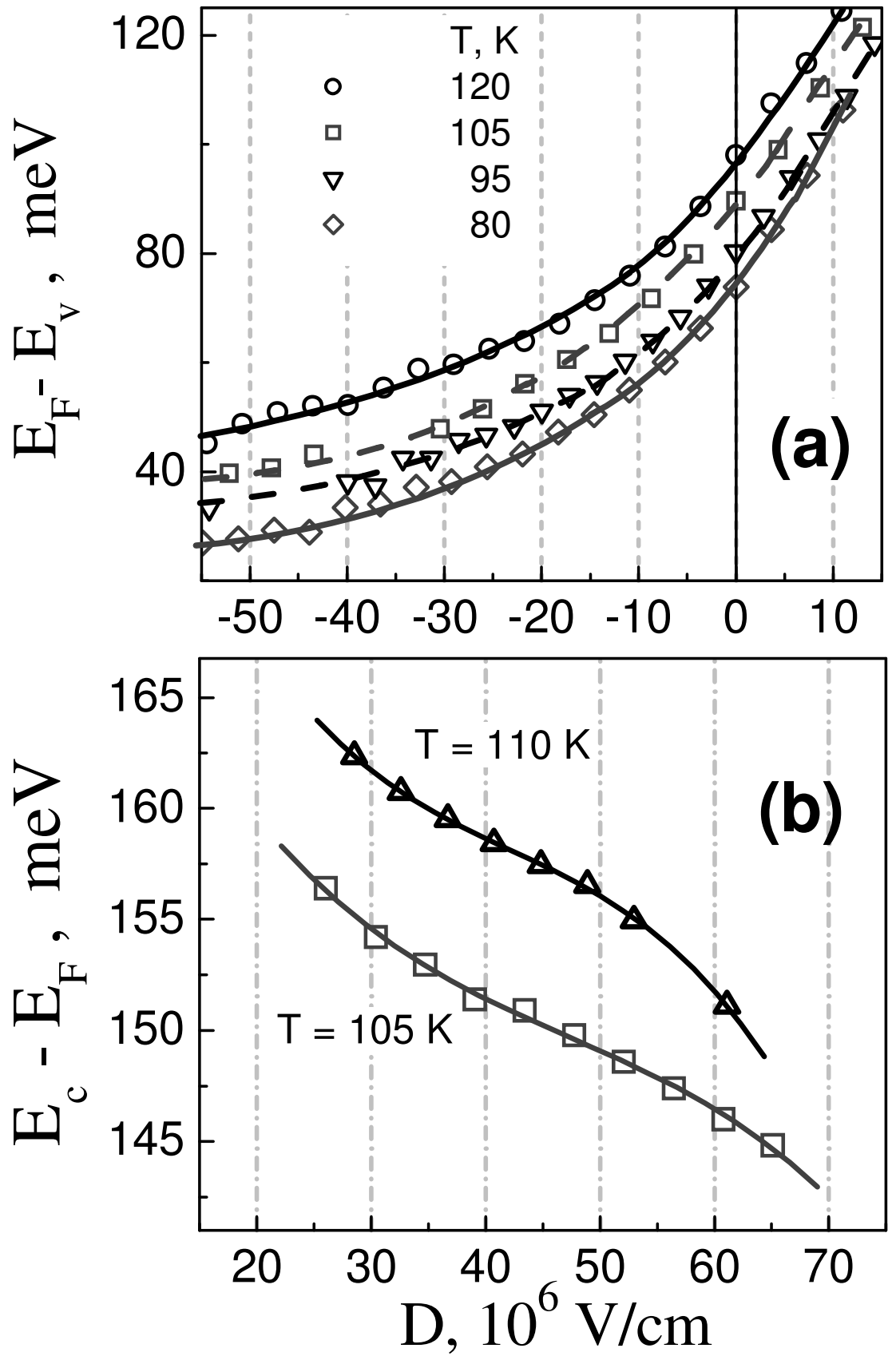

Fig. 9

Butenko et al

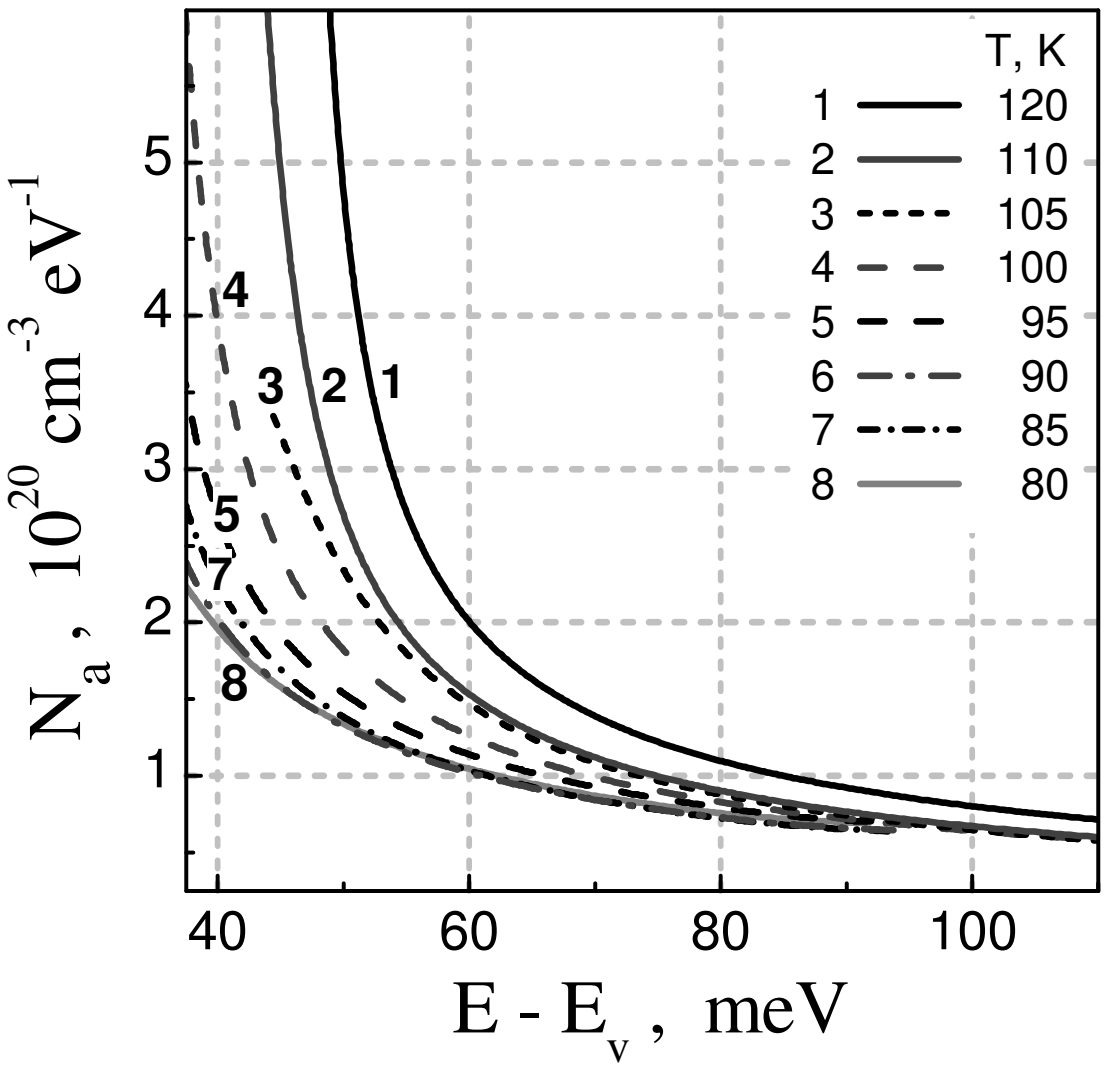

**Fig. 10.**

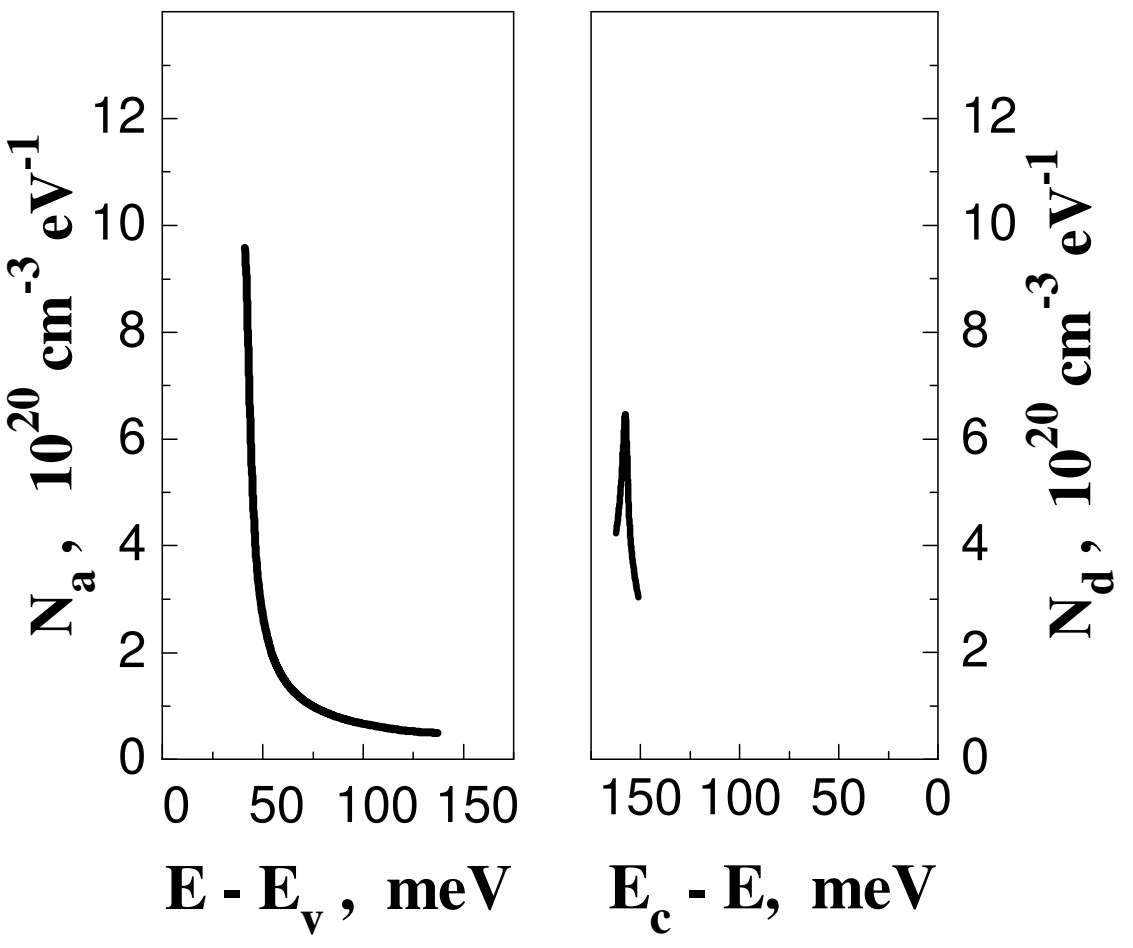

$N_a$, $10^{20}$ cm$^{-3}$ eV$^{-1}$
0
2
4
6
8
10
12
0
50
100
150
E - $E_v$, meV
$N_d$, $10^{20}$ cm$^{-3}$ eV$^{-1}$
150
100
50
0
$E_c$ - E, meV

**Fig. 11.**

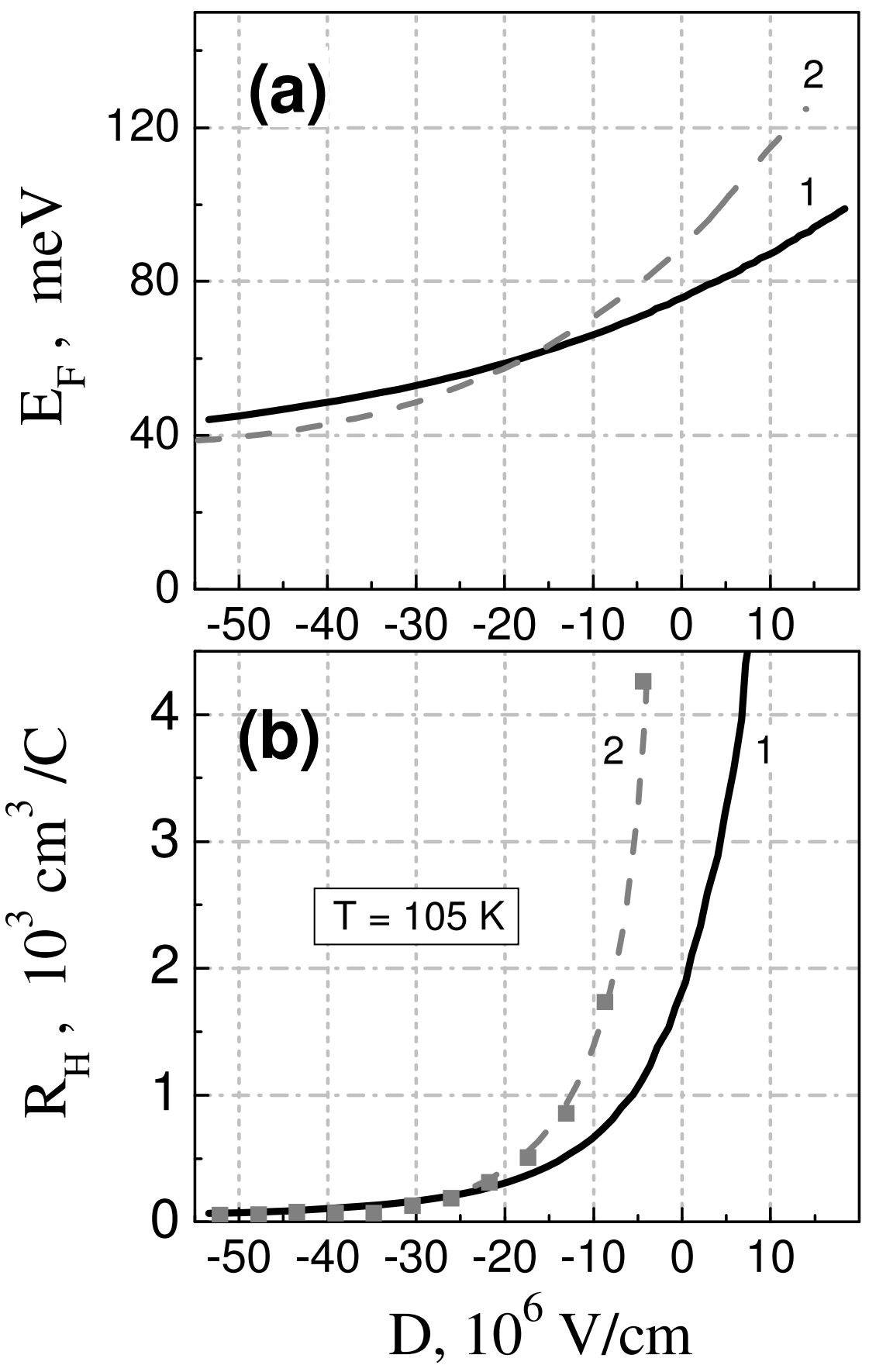

**Fig. 12.**